\documentclass[journal]{IEEEtran}
\usepackage{cite}
\usepackage{balance}
\usepackage{makecell}

\ifCLASSINFOpdf
   \usepackage[pdftex]{graphicx}
   \usepackage{caption}
\else
\fi

\begin{document}
%
\title{Microprocessor Optimizations for the Internet of Things: A Survey}
%
\author{Tosiron~Adegbija,~\IEEEmembership{Member,~IEEE,}
        Anita~Rogacs,
        Chandrakant~Patel,~\IEEEmembership{Fellow,~IEEE,}
        and~Ann~Gordon-Ross,~\IEEEmembership{Member,~IEEE,}

\thanks{T. Adegbija is with the Department
	of Electrical and Computer Engineering, University of Arizona, USA, e-mail: tosiron@email.arizona.edu.}
\thanks{A. Rogacs and C. Patel are with Hewlett-Packard (HP) Labs, USA, e-mail: rogacs@hp.com, chandrakant.patel@hp.com.}
\thanks{A. Gordon-Ross is with the University of Florida, USA and the Center for High Performance Reconfigurable Computing (CHRE) at UF. e-mail: ann@ece.ufl.edu.}

}


\maketitle

\begin{abstract}
The Internet of Things (IoT) refers to a pervasive presence of interconnected and uniquely identifiable physical devices. These devices' goal is to gather data and drive actions in order to improve productivity, and ultimately reduce or eliminate reliance on human intervention for data acquisition, interpretation and use.  The proliferation of these connected low-power devices will result in a data explosion that will significantly increase data transmission costs with respect to energy consumption and latency. \textit{Edge computing} reduces these costs by performing computations at the edge nodes, prior to data transmission, to interpret and/or utilize the data. While much research has focused on the IoT's connected nature and communication challenges, the challenges of IoT embedded computing with respect to device microprocessors has received much less attention. This article explores IoT applications' execution characteristics from a microarchitectural perspective and the microarchitectural characteristics that will enable efficient and effective edge computing. To tractably represent a wide variety of next-generation IoT applications, we present a broad IoT application classification methodology based on application functions, to enable quicker workload characterizations for IoT microprocessors. We then survey and discuss potential microarchitectural optimizations and computing paradigms that will enable the design of right-provisioned microprocessors that are efficient, configurable, extensible, and scalable. Our work provides a foundation for the analysis and design of a diverse set of microprocessor architectures for next-generation IoT devices.
\end{abstract}

\begin{IEEEkeywords}
Internet of Things, edge computing, low-power embedded systems, microprocessor optimizations, IoT survey, adaptable microprocessors, heterogeneous architectures, energy harvesting, approximate computing.
\end{IEEEkeywords}

%
\IEEEpeerreviewmaketitle

\section{Introduction and Motivation}

\IEEEPARstart{T}{he} Internet of Things (IoT) is an emerging technology that refers to a pervasive presence of interconnected and uniquely identifiable physical devices, comprising an expansive variety of devices, protocols, domains, and applications. The IoT will involve devices that gather data and drive actions in order to improve productivity, and ultimately reduce or eliminate reliance on human intervention for data acquisition, interpretation, and use \cite{ashton09}. The IoT has been described as one of the disruptive technologies that will transform life, business, and the global economy \cite{mckinsey}. Based on analysis of key potential IoT use-cases (e.g., healthcare, smart cities, smart home, transportation, manufacturing, etc.), it has been estimated that by 2020, the IoT will constitute a trillion dollar economic impact and include more than 50 billion low-power devices that will generate petabytes of data \cite{gartner,sundmaeker10,zhou11}.

\begin{figure}
	\centerline{\includegraphics[width=0.5\textwidth]{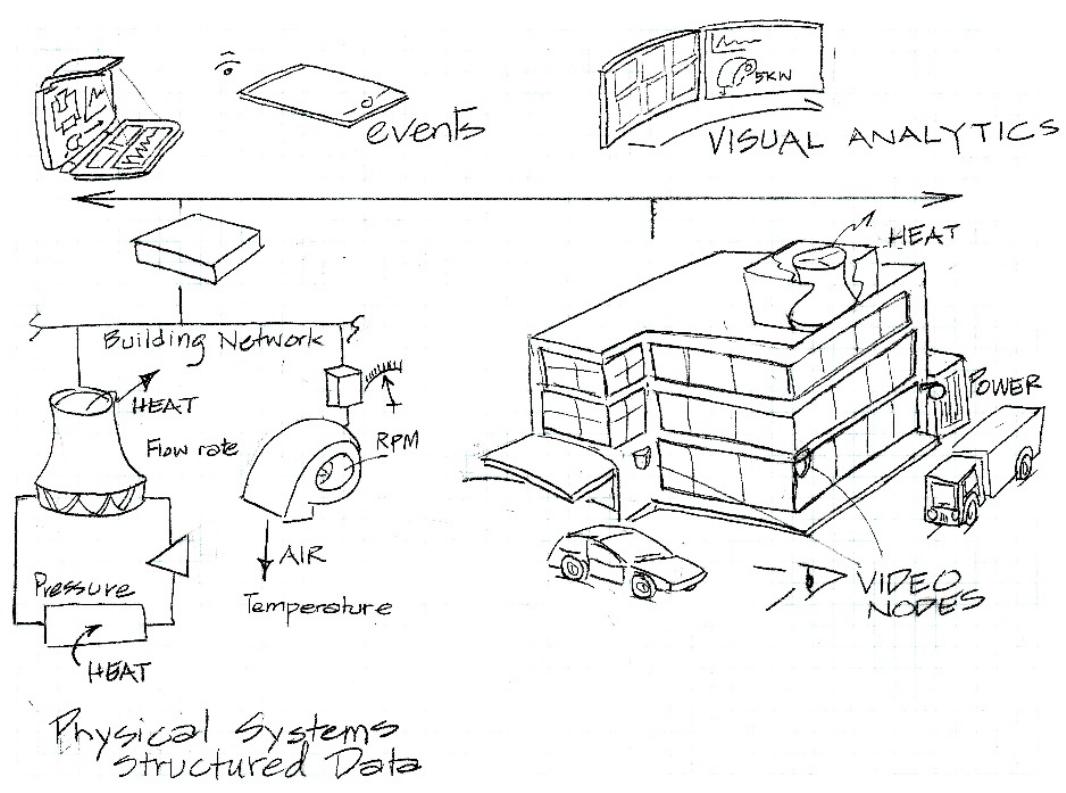}}
	\caption{Illustration of the high-level components of the Internet of Things.}
	\label{fig1}
\end{figure}

Due to the IoT's expected growth and potential impact, much research has focused on the IoT’s communication and software layer \cite{atzori,gubbi13,miorandi12,li11}, however, the challenges of IoT computing, especially with respect to device microprocessors, has received much less attention. Computing on IoT devices introduces new substantial challenges, since IoT devices' microprocessors must satisfy increasingly growing computational and memory demands, maintain connectivity, and adhere to stringent design and operational constraints, such as low cost, low energy budgets, and in some cases, real-time constraints. These challenges necessitate new research focus on microarchitectural optimizations that will enable designers to develop right-provisioned architectures that are efficient, configurable, extensible, and scalable for next-generation IoT devices.

Figure \ref{fig1} depicts an IoT use-case that illustrates the high-level components of the traditional IoT model. The IoT typically comprises of several low-power/low-performance edge nodes, such as sensor nodes, that gather data and transmit the data to high-performance head nodes, such as servers, that perform computations for visualization and analytics. In a data center, for example, data aggregation from edge nodes facilitates power and cooling management\cite{bash06,patel03}.

However, the growth of the IoT and the resulting exponential increase in acquired/transmitted data poses significant bandwidth and latency challenges. These challenges are exacerbated by the intrinsic resource constraints of most embedded edge nodes (e.g., size, battery capacity, real-time deadlines, cost, etc.). These resource constraints must be taken into account in the design process, and may make it more difficult to achieve design objectives (e.g., minimizing energy, size, etc.). Additionally, increasing consumer demands for high-performance IoT applications will necessitate acquisition and transmission of complex data. For example, a potentially impactful IoT use-case is medical diagnostics \cite{medical}. With the advent of technological advances such as cheap portable magnetic resonance imaging (MRI) devices and portable ultra-sound machines, several gigabytes (GBs) of high resolution images will be transmitted to medical personnel for remote data processing and medical diagnosis. In some cases, this system must scale to a network of several portable medical devices that transfer data to medical personnel. Transmitting this data will result in bandwidth bottlenecks and pose additional challenges for real-time scenarios (e.g., medical emergencies) where the latency must adhere to stringent deadline constraints.

The IoT can also incur significant, and potentially unsustainable, energy overheads. Previous work \cite{kwok06,baliga11} established that energy consumed while transmitting data is significantly more than the energy consumed while performing computations on the data. For example, the energy required by Rockwell Automation’s sensor nodes to transmit one bit of data is 1500-2000X more than the energy required to execute a single instruction (depending on the transmission range and specific computations) \cite{raghunathan04}. 

To address these challenges, \textit{fog computing} \cite{bonomi12} has been proposed as a virtualized platform that provides compute, storage, and networking services between edge nodes and cloud computing data centers. Rather than performing computations in the cloud, fog computing reduces the bandwidth bottleneck and latency by moving computation closer to the edge nodes. Our study focuses on further reducing the bandwidth, latency, and energy consumption through \textit{edge computing}, where the edge nodes are directly equipped with sufficient computation capacity in order to minimize data transmission \cite{adegbija15_iot}.  

Edge computing performs computations that process, interpret, and use data at the edge nodes. Performing these computations on the edge nodes minimizes data transmission, thereby improving latency, bandwidth, and energy consumption. For example, in the medical diagnostics use-case described above, rather than sending several GBs of MRI data to the medical personnel for diagnoses, the portable MRI machine (the edge node) is equipped with sufficient computational capabilities and algorithms to extract information and interpret the data. Only processed data (e.g., information about an anomaly in the patient) is transmitted to the medical personnel, thus speeding up the diagnoses and reducing the MRI machine's energy consumption. Alternatively, the data could be quantifiably reduced using intelligent algorithms and computations, such that only important information is transmitted to the medical personnel. 

Gaura et al. \cite{gaura13} examined the benefits of edge mining, in which data mining takes place on the edge devices. The authors showed that edge mining has the potential to reduce the amount of transmitted data, thus reducing energy consumption and storage requirements. However, the edge nodes’ computing capabilities must be sufficient/right-provisioned to perform and sustain the required computations, while adhering to the nodes’ design constraints (e.g., form factor, energy consumption, etc.) \cite{sanchez14}. 

This paper explores microarchitectural optimizations and emerging computing paradigms that will enable edge computing on the IoT. To ensure that microprocessor architectures designed and/or selected for the IoT have sufficient computing capabilities, a holistic approach, involving both application and microarchitecture characteristics, must be taken to determine microarchitectural design tradeoffs. However, due to the wide variety of IoT applications and the diverse set of available architectures, determining the appropriate architectures is very challenging. The study presented herein seeks to address these challenges and motivate future research in this direction.

In this paper, we perform an expansive study and characterization of the emerging IoT application space and propose an application classification to broadly represent IoT applications with respect to their execution characteristics. To enable the design of right-provisioned microprocessors, we propose the use of computational kernels that provide a tractable starting point for representing key computations that occur in the IoT application space. Using computational kernels, rather than full applications, follows the computational dwarfs methodology \cite{asanovic06} and allows IoT computational patterns to be accurately represented at a high level of abstraction. 

Furthermore, we propose a high-level design methodology for identifying right-provisioned architectures for edge computing use-cases, based on the executing applications and the applications’ execution characteristics (e.g., compute intensity, memory intensity, etc.). Finally, in order to motivate future research, we survey a few potential microprocessor optimizations and computing paradigms that will enable the design of right-provisioned IoT microprocessor architectures.

\section{High-level IoT Characteristics and their Demands on Microprocessor Architectures}

The IoT's characteristics necessitate new designs and optimizations for microprocessors that will be employed in IoT devices. We briefly describe seven key characteristics---based on previous research \cite{perera14}---that, together, distinguish the IoT from other connected systems: \textit{intelligence, heterogeneity, complexity, scale, real-time constraints, spatial constraints,} and \textit{inter-node support}. We also describe the demands that these characteristics place on microprocessor architectures

\begin{itemize}
	\item \textbf{Intelligence:} Since the goal of the IoT is to reduce reliance on human intervention for data acquisition and use \cite{ashton09}, raw data must be autonomously collected and processed to create actionable information. IoT microprocessors must be able to dynamically adapt to varying runtime execution scenarios and adaptable data characteristics \cite{gopinath11}.
	
	\item \textbf{Heterogeneity:} One of the key characteristics of the IoT is that it involves a high degree of heterogeneity, featuring different kinds of devices, applications, and contexts \cite{perera14}. Thus, IoT microprocessors must be specialized to the different execution characteristics of IoT applications. IoT microprocessor heterogeneity may be \textit{chip-level}---a single chip with heterogeneous cores---or \textit{network-level}, where different devices feature different kinds of cores. Despite this heterogeneity, the devices must be able to seamlessly communicate with each other and share resources for efficient data interpretation and use.
	 
	\item \textbf{Complexity:} The organization and management of the IoT will be very complex. Apart from the large numbers of heterogeneous architectures, the architectures must be able to execute a wide variety of applications, many of which may be memory- and compute-intensive. Interactions between the different IoT devices will dynamically vary. Some devices will be added to the IoT network, while others will be removed; these changes may impact individual devices' execution behaviors.
	
	\item \textbf{Scale:} The IoT will comprise more than 50 billion devices by 2020, and the numbers are expected to grow continuously \cite{sundmaeker10}. In addition to the increase in the number of devices, the interactions among them will also increase. To support this scale, IoT microprocessors must be efficient---cost, energy, and area efficient---and constitute minimal overhead to the IoT device. In addition, the microprocessors must be able to portably execute different kinds of applications.
	 
	\item \textbf{Real-time constraints:} Some of the most important IoT use-cases---for example, patient monitoring, medical diagnostics, aircraft monitoring---involve real-time constraints, where execution must adhere to stringent deadlines. IoT microprocessors must be able to dynamically determine and adhere to deadlines, based on various inputs, such as user inputs, application characteristics, quality of service.
	
	\item \textbf{Spatial constraints:} Several IoT use-cases are location-based. An IoT device's location may change throughout the device's lifetime. In addition, the device may be exposed to variable, and potentially non-ideal, environmental conditions. For example, tracking devices may be exposed to extreme heat, extreme cold, and/or rain at different times or in different locations. Thus, IoT microprocessors must feature fault tolerance and adaptability that allows them to adhere to variable operation conditions. 
	
	\item \textbf{Inter-node support:} The IoT will comprise of several devices/nodes that can share execution resources among each other. Due to the wide variety of IoT applications that may execute on a device, and the stringent resource constraints, it may be impractical to equip every device with all the execution resources it will require throughout its lifetime. Thus, to maintain efficient execution, IoT devices must be able to share execution resources with each other, when necessary.
	
\end{itemize}

\section{IoT Application Classification} \label{classes}
The IoT offers computing potential for many application domains, including transportation and logistics, healthcare, smart environment, personal and social domains \cite{atzori}, etc. One of the key goals of the IoT, from an edge computing perspective, is to equip edge devices with sufficient resources to perform computations that would otherwise have been transferred to a high-performance device. In order to rightly provision these devices, we must first understand potential applications that will be executed on the devices. 

Previous works have proposed classifications for various IoT components. Gubbi et al. \cite{gubbi13} presented a taxonomy for a high level definition of IoT components with respect to hardware, middleware, and presentation/data visualization. Tilak et al. \cite{tilak02} presented a taxonomy to classify wireless sensor networks according to different communication functions, data delivery models, and network dynamics. Tory et al. \cite{tory04} presented a high level visualization taxonomy that classified algorithms based on the characteristics of the data models. 

However, there is currently very little research that characterizes these applications with respect to their execution characteristics. One of the biggest challenges the IoT presents is the huge number and diversity of use-cases and potential applications that will be executed on IoT devices. This challenge is exacerbated by the fact that only a small fraction of these applications are currently available in society. Thus, a significant amount of foresight is required in designing microprocessor architectures to support the IoT's emergence and growth.

Much prior work has characterized IoT applications according to different use-cases and domains. For example, Atzori et al. \cite{atzori} and Sundmaeker et al. \cite{sundmaeker10} categorized IoT applications into three domains: industry, environment, and society. Asin et al. \cite{asin12} categorized IoT applications into 54 domains under twelve categories. In this work, our goal is a tractable and extensible classification that enables us to identify the IoT applications' key execution characteristics.

As an initial step towards understanding IoT applications' execution characteristics, we performed an expansive study of IoT use-cases and the application functions present in these use-cases. Since it is impractical to consider every IoT application within these use-cases/application domains, based on our study, we propose an application classification methodology that provides a high level, broad, and tractable representation of a variety IoT applications using the application functions. Our IoT application classification consists of six key application functions: 

\begin{itemize}
	\item \textit{sensing} 
	\item \textit{communications}
	\item \textit{image processing} 
	\item \textit{compression (lossy/lossless)} 
	\item \textit{security}
	\item \textit{fault tolerance}.
\end{itemize}

 We note that this classification is not exhaustive; however, it represents a wide variety of current and potential IoT applications. The classification also provides an extensible framework that allows emerging applications/application domains to be analyzed. In this section, we describe the application functions and motivate these functions using a medical diagnostics use-case, where applicable, or other specific examples of current and/or emerging IoT applications.

\subsection{Sensing}

Sensing involves data acquisition (e.g., temperature, pressure, motion, etc.) about objects or phenomena, and will remain one of the most common functions in IoT applications. In these applications, activities, information, and data of interest are gathered for further processing and decision making. We use sensing in our IoT application classification to represent applications where data acquired using sensors must be converted to a more useable form. Our motivating example for sensing applications is \textit{sensor fusion} \cite{nakamura07}, where sensed data from multiple sensors are fused to create data that is considered qualitatively or quantitatively more accurate and robust than the original data. 

Sensor fusion algorithms can involve various levels of complexity and compute/memory intensity. For example, sensor fusion could involve aggregating data from various sources using simple mathematical computations, such as addition, minimum, maximum, mean, etc. Alternatively, sensor fusion could involve more computationally complex/expensive applications, such as fusing vector data (e.g., video streams from multiple sources), which requires a substantial increase in intermediate processing. 

In a medical diagnostics use-case, for example, sensing is vital in a body area network \cite{chen11}, where non-invasive sensors can be used to automatically monitor a patient’s physiological activities, including blood pressure, heart rate, motion, etc. Several sensing devices, such as portable electrocardiography (ECG), electroencephalography (EEG), and electromyography (EMG) machines, motion and blood pressure sensors could be equipped with additional computational resources and algorithms that enable the devices to not only gather data, but also analyze the data in order to reduce the amount of transmitted data, with minimal energy or area overheads.

\subsection{Communications}

Communications is one of the most common IoT application functions due to the IoT’s intrinsic connected structure, where data transfers traverse several connected nodes. There are many communication technologies (e.g., Bluetooth, Wi-Fi, etc.), and communication protocols (e.g., transfer control protocol (TCP), the emerging 6lowpan (IPv6 over low power wireless personal area network), etc.). In this work, we highlight \textit{software defined radio (SDR)} \cite{lee05}, which is a communication system in which physical layer functions (e.g., filters, modems, etc.) that are typically implemented in hardware are implemented in software.

SDR is an emerging and rapidly developing communication system that is driving the innovation of communications technology, and promises to impact all areas of communication. SDR is growing in popularity, and attractive for the IoT, because of its inherent flexibility, which allows for flexible incorporation and enhancements of multiple radio functions, bands, and modes, without requiring hardware updates. SDR typically involves an antenna, an analog-to-digital converter (ADC) connected to an antenna (for receiving) and a digital to analog converter (DAC) connected to the antenna (for transmitting). Digital signal processing (DSP) operations (e.g., Fast Fourier Transform (FFT)) are then used to convert the input signals to any form required by the application. 

Even though SDR applications are typically compute intensive, with small data and instruction memory footprints, recent work \cite{chen16} shows that the overheads of SDR can be kept small in the IoT domain by focusing on optimizing the key kernels (e.g., Synchronization and Finite Impulse Response (FIR)) that dominate SDR computations and power consumption. In general, SDR algorithms can be efficiently executed using general purpose microprocessors or more specialized processors, such as digital signal processors (DSPs) or field-programmable gate arrays (FPGAs). Alternatively, heterogeneous architectures \cite{jeff12} can also combine different kinds of microprocessors to satisfy different operations' execution requirements while minimizing overheads, such as energy consumption. Other examples of communication applications include packet switching and TCP/IP.

\subsection{Image Processing}

In the IoT context, image processing represents applications that involve any form of signal processing where the input is an image or video stream from which characteristics/parameters must be extracted/identified. Additionally, this classification also involves applications in which an image/video input must be converted to a more usable form. Several emerging IoT applications, such as automatic number license plate recognition, traffic sign recognition, face recognition, etc., involve various forms of image processing. For example, face recognition involves operations, such as face detection, landmark recognition, feature extraction, and feature classification, all of which involve image processing. 

Image processing is important for several impactful IoT use-cases, and necessitates microarchitectures that can efficiently perform image processing operations. For example, in medical diagnostics, image processing can be used to increase the reliability and reproducibility of disease diagnostics. Image processing can provide medical personnel with quantitative data from historical images, which can be used to supplement qualitative data currently used by specialists. In addition, portable medical devices, e.g., portable ultrasounds, can be equipped with image processing applications to provide speedy analysis for remote assessment of patients \cite{mort16}. 

The National Institute of Health (NIH) supports the Medical Image Processing, Analysis, and Visualization (MIPAV) application \cite{medical}, which enables medical researchers to easily share research data and enhance their ability to diagnose, monitor, and treat medical disorders. However, since image processing applications are typically data-rich, and both memory and compute intensive, novel optimization techniques are required to enable the efficient execution of these applications in the context of IoT edge computing. Furthermore, some image processing applications require large input, intermediate, or output data to be stored (e.g., medical imaging), thus requiring a large amount of storage.

\subsection{Compression}

With the increase in data and bandwidth-limited systems, compression can reduce communication requirements to ensure that data is quickly retrieved, transmitted, and/or analyzed. Several emerging IoT use-cases will involve large volumes of data, which will necessitate efficient compression techniques to accommodate the rapid growth of the data and reduce transmission latency and bandwidth costs \cite{xiong03}. Additionally, since most IoT devices are resource-constrained, compression also reduces storage requirements when data must be stored on the edge node. For example, data gathered using sensors in a body area network can be quantifiably and intelligently reduced in order to minimize transmission and storage requirements for medical diagnosis devices.

Compression involves encoding information using fewer bits than the original representation. The data can be encoded at the data source before storage or transmission, known as source encoding, or during transmission, known as channel coding \cite{anderson12}. In our studies, however, we focus on source encoding, as this type of encoding will be more relevant in the context of edge computing. 

Compression techniques can be broadly classified as \textit{lossy} or \textit{lossless} compression. Lossy compression (e.g., JPEG) typically exploits the perceptibility of the data in question, and removes unnecessary data, such that the lost data is imperceptible to the user. Alternatively, lossless compression removes statistically redundant data in order to concisely represent data. Lossless compression typically achieves a lower compression ratio and is usually more compute and memory intensive than lossy compression. However, lossy compression may be unsuitable in some scenarios where high data fidelity is required to maintain the quality of service (QoS) (e.g., in medical imaging).

\subsection{Security}

Since IoT devices are often deployed in open or potentially unsafe environments, where the devices are susceptible to malicious attacks, security applications are necessary to maintain the integrity of both the devices and the data. Furthermore, sensitive scenarios (e.g., medical diagnostics) may require security applications to prevent unauthorized access to sensitive data and functions. Implantable medical devices, such as pacemakers, implantable cardiac defibrillators, neurostimulators are especially susceptible to potentially fatal security and privacy issues, such as replay attacks \cite{halperin08,karimian16}. Since medical device security is still in its infancy, there still exists a wide knowledge gap with respect to the microprocessor characteristics that will support security algorithms’ execution requirements without sacrificing the devices’ functional requirements.

We highlight data encryption \cite{singh11}, which is a common technique for ensuring data confidentiality, wherein an encryption algorithm is used to generate encrypted data that can only be read/used if decrypted. Data encryption applications (e.g., secure hash algorithm) are typically compute intensive and memory intensive, since encryption speed is also dependent on the memory access latency for data retrieval and storage.

\subsection{Fault Tolerance}

Fault tolerance \cite{izosimov05} refers to a system's ability to operate properly when some of its components fail. Fault tolerant applications are especially vital since IoT devices may be deployed in harsh and unattended environments, where QoS must be maintained in potentially adverse conditions, such as cryogenic to extremely high temperatures, shock, vibration, etc. In some emerging IoT devices, such as implantable medical devices, fault tolerance could be the single most critical requirement, since faults can be potentially fatal. Thus, fault tolerance must be incorporated into such devices without accruing significant overheads.

Fault tolerance can be achieved in different ways. Hardware-based techniques usually rely on redundancy---RAID (redundant array of independent disks) \cite{patterson88} is a common example---wherein redundant disks or devices are used to provide fault tolerance in the event of a failure. This kind of redundancy can be achieved in IoT devices using a dedicated IoT device, or integrated into a larger, less constrained device, in order to minimize the attendant overheads of redundancy. Alternatively, redundancy can be incorporated directly into the IoT devices, at the expense of area and power overheads. To reduce the overheads from hardware-based fault tolerance, software-based fault tolerance  can also be employed. Software-based fault tolerance \cite{izosimov05,unsal02,saha06} involves applications and algorithms that perform operations, such as memory scrubbing, cyclic-redundancy checks (CRC), error detection and correction, etc.

\begin{figure}[t]
	\centering
	\captionsetup{justification=centering}
	\includegraphics{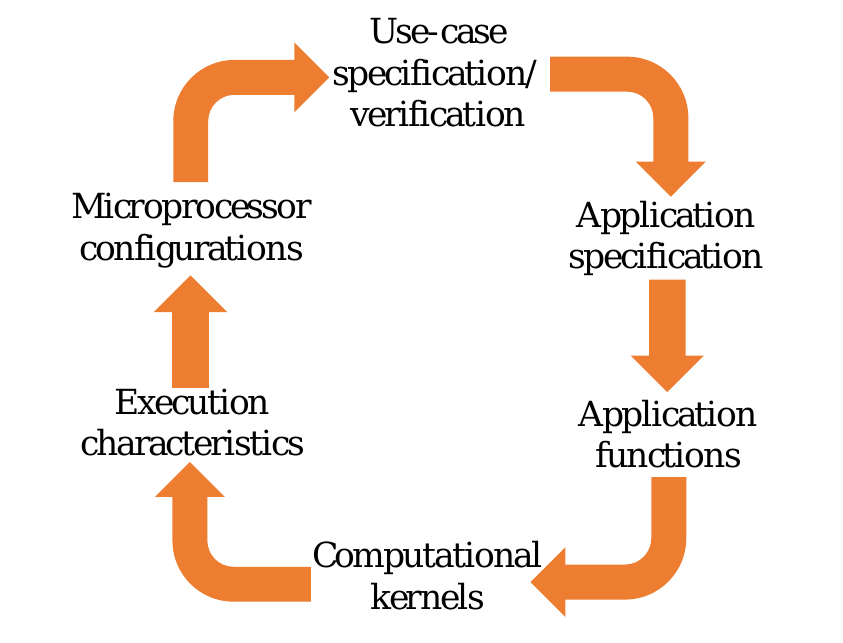}
	\caption{Illustration of a high-level IoT microprocessor design life-cycle}
	\label{fig:lifecycle}
\end{figure}

\section{Determining IoT Microprocessor Configurations}

One of the major challenges for IoT microprocessor design is determining the best microprocessor configurations that satisfy the IoT device's execution requirements. In this section, we describe a sample high-level process through which an IoT microprocessor can be designed and optimized. Figure \ref{fig:lifecycle} illustrates a high-level IoT microprocessor design life-cycle, consisting of six steps. First, the use-case needs to be specified. This step describes the overall functionality and behavior of the IoT device, which will dictate the microprocessor requirements. Based on the use-case, the required applications to achieve the desired functionality are then specified. For example, a medical diagnostics use-case involving a portable ultrasound device \cite{kim12} may require applications for image capture, anomaly detection, anomaly recognition, data encryption, and data transmission. Thereafter, the specific functions within each application are determined, and these functions are broken down into their respective \textit{computational kernels}. 

Computational kernels are basic execution blocks that represent applications' functions; the kernels disconnect the executions from specific implementations, programming languages, and algorithms. Using computational kernels can make the design process more manageable, since kernels are faster to simulate and can represent a variety of applications/application functions. In addition, kernels expose computational nuances and reveal execution characteristics that may not be visible when considering the full application. Kernels also provide a fine-grained view of applications, such that additional design processes (e.g., hardware/software partitioning \cite{stitt03}) can be sped up. Using computational kernels to represent application functions is supported by the concept of \textit{computational dwarfs} \cite{asanovic06}. Computational dwarfs represent patterns of computation at high levels of abstraction to encompass several computational methods in modern computing. 

\begin{table} 
	\begin{center}
\captionof{table}{Application functions and sample representative kernels.}
	\begin{tabular}{| l || c |} 
	\hline \label{table:kernels}
	\textbf{Application function} & \textbf{Kernel} \\ \hline
	Sensing & dense matrix transpose \\ \hline
	Communications & Fast Fourier Transform (FFT) \\ \hline
	Image processing & Dense matrix multiplication \\ \hline
	Lossy compression & jpeg \\ \hline
	Lossless compression & lz4 \\ \hline
	Security & Secure Hash Algorithm (sha)\\ \hline
	Fault tolerance & Cyclic redundancy check (crc) \\ 
	\hline
\end{tabular}
	\vspace{-15pt}
\end{center}
\end{table}

Table \ref{table:kernels} illustrates sample kernels that can be used to represent the different application classes (Section \ref{classes}). After the computational kernels are determined, the kernels' execution characteristics are then determined. One of the most common way for determining these characteristics is through simulations. During design space exploration \cite{shao14}, the kernels' execution characteristics---memory intensity, compute intensity, instructions per cycle, memory references, etc.---using different microprocessor configurations are then analyzed to determine which configurations best satisfy the application resource requirements. These configurations can then be refined, if necessary, after verifying that the functional requirements of the use-case are met.

\section{State of the Art in IoT Microarchitecture Configurations} \label{sec:configs}

We performed an extensive survey and study of the state-of-the-art in commercial-off-the-shelf (COTS) embedded systems microprocessor architectures from several designers and manufacturers ranging from low-end microcontrollers to high-end/high-performance low-power embedded systems microprocessors. Our studies included publicly available information on these microprocessors' configurations, and conversations with researchers and engineers directly involved with microprocessor design and development in several manufacturing companies. 

Based on our studies, we categorized the microprocessors in terms of several microprocessor characteristics, including number of cores, on-chip memory (e.g., cache), off-chip memory support, power consumption, number of pipeline stages, etc. Using this information, we developed a set of four high-level microarchitecture configurations for IoT edge computing support. These configurations represent the range of available state-of-the-art COTS microprocessors, and provide a reference point from which future IoT microprocessors and optimizations can be developed. While microprocessors could include central processing units (CPUs), graphics processing units (GPUs), DSPs, etc., in this survey, we focus on CPUs, since they are typically the backbone for most edge computing applications.

\begin{table*}
	\begin{center}
	\captionof{table}{State of the art microprocessor configurations.}
	\begin{tabular}{| l || c | c | c | c |}
		\hline \label{table:configs}
		 & \textbf{Config1} & \textbf{Config2} & \textbf{Config3} & \textbf{Config4} \\ \hline
		\textbf{Sample CPU} & ARM Cortex M4 & Intel Quark & ARM Cortex A7 & ARM Cortex A15 \\ \hline
		\textbf{Frequency} & 48 MHz & 400 MHz & 1 GHz & 1.9 GHz \\ \hline
		\textbf{Number of cores} & 1 & 1 & 4 & 4 \\ \hline
		\textbf{Pipeline stages} & 3 & 5 & 8 & 15 \\ \hline
		\textbf{Cache} & None & None & 32KB I/D L1, 1MB L2 & 32KB I/D L1, 2MB L2 \\ \hline
		\textbf{Memory} & 512KB flash & 2GB RAM & 2GB RAM support & 1TB RAM support\\ \hline
		\textbf{Execution} & In-order & In-order & In-order & Out-of-order \\ 
		\hline
	\end{tabular}
	\end{center}
\end{table*}

Table \ref{table:configs} depicts the microarchitecture configurations, comprising of four configurations: \textit{config1, config2, config3,} and \textit{conf4}, representing different kinds of microprocessors. We highlight specific state-of-the-art microcontroller/microprocessor examples to motivate the configurations, however, we note that these configurations are only representative and not necessarily descriptive.

\textit{Config1} represents low-power and low-performance microcontroller units, such as the ARM Cortex-M4 \cite{cortexM4} found in several IoT-targeted MCUs from several developers, including Freescale Semiconductors, Atmel, and STMicroelectronics. Conf1 contains a single core with 48 MHz clock frequency, three pipeline stages, in-order execution, and support for 1 MB of flash memory. 

\textit{Config2} represents recently-developed IoT-targeted CPUs, such as the Intel Quark Technology \cite{intelQuark}, and contains a single core with 400 MHz clock frequency, five pipeline stages, in-order execution, 16 KB level one (L1) instruction  and data caches, and support for 2 GB RAM.

\textit{Config3} represents mid-range CPUs, such as the ARM Cortex-A7 \cite{cortexA7_A15} found in several general purpose embedded systems, and contains four cores with 1 GHz clock frequency, 8 pipeline stages, in-order execution, 32 KB L1 instruction and data caches, 1 MB level two (L2) cache, and support for 2 GB RAM. 

Finally, \textit{config4} represents high-end/high-performance embedded systems CPUs, such as the ARM Cortex-A15 \cite{cortexA7_A15}, and contains four cores with 1.9 GHz clock frequency, 8 pipeline stages, 32 KB L1 instruction and data caches, 2 MB L2 cache, support for 4 GB RAM, and out-of-order execution. Out-of-order execution allows instructions to execute as soon as the instruction becomes available, unlike in-order execution where instructions must execute in program order.

\section{Microprocessor Optimizations, Computing Paradigms, and Future Directions}

Using the configurations described in Section \ref{sec:configs} and the kernels listed in Table \ref{table:kernels} as benchmarks, we performed detailed architectural simulations on GEM5 \cite{gem5} to analyze the characteristics of the configurations. The details of our analysis can be found in our preliminary work \cite{adegbija15_iot}. Based on our analysis and extensive surveys, we have identified five key characteristics that IoT microprocessors must have in order to support the IoT's growth:

\begin{itemize}
	\item \textbf{Efficiency}: Due to the typical stringent resource constraints of IoT devices, IoT microprocessors must be optimized for energy, cost, performance, and area efficiency.
	\item \textbf{Configurability}: One of the major observations from our studies is that different IoT applications have vastly different runtime resource requirements. With the growth of the IoT and the current trend of IoT applications, we envision that this variability in runtime resource requirements will increase even further with next-generation IoT applications. Therefore, these variable runtime resource requirements necessitate adaptable/configurable microarchitectures with configurations that can be autonomously specialized to different applications in order to achieve optimal execution, especially in terms of energy efficiency.
	\item \textbf{Security}: Security must be one of the primary design goals of IoT microprocessors, especially since IoT devices will be inherently more susceptible to attacks.
	\item \textbf{Future-proof}: Designing IoT microprocessors will require a lot of foresight. With the rapid emergence of new IoT applications, and the applications' increasing memory and compute requirements, IoT microprocessors must be able to execute future applications without being over-provisioned for current applications.
	\item \textbf{Extensibility}: Future-proofing IoT microprocessors can be achieved by extending the microprocessors with additional functionalities (e.g., specialized instructions, security monitors, new on-chip peripherals). Thus, IoT microprocessors must be designed with ease of integration, customization, and extension in mind. Such extensibility will allow new levels of performance and energy efficiency to be achieved.
\end{itemize}

In this section, we survey a few potential microprocessor optimizations and computing paradigms, from a context of edge computing, that will enable the aforementioned IoT characteristics, and support the IoT's growth.  We first discuss optimizations for achieving adaptability in IoT microprocessors; we then survey and discuss research directions in other computing paradigms that will enable microprocessor optimizations for the IoT, including \textit{non-volatile processors}, \textit{approximate computing}, \textit{in-memory processing}, and \textit{secure microarchitectures}. Our goal in this section is not to provide an exhaustive survey of potential microprocessor optimizations; our goal is to motivate researchers with future directions for developing novel, configurable, and extensible low-overhead IoT microprocessors. Table \ref{table:paradigms} summarizes the different computing paradigms, benefits, and references for further details.

\subsection{Configurable/Adaptable Architectures}

In order to achieve the right balance between performance, power, and area in IoT applications, IoT microprocessors must be adaptable to the application requirements. This adaptability can be achieved through the ability to change the microprocessor's configuration at runtime or by heterogeneous processors that offer different processing resources for executing the applications. Several microprocessor components can be configured, including the issue queue \cite{folegnani01}, reorder buffer \cite{kora13}, register files \cite{abella03}, and pipelines \cite{efthymiou03}. However, in this subsection, we focus on configurable caches \cite{zhang03} due to their potential impact on the microprocessor's end-to-end energy, performance, and area.

The memory will arguably remain the most important microprocessor component for performance and energy consumption. Since emerging IoT applications will increase in memory and compute intensity, IoT microprocessors must be equipped with more advanced memory hierarchies to take advantage of the spatial and temporal locality of the IoT applications. Due to the memory hierarchy's large impact on system performance and energy consumption, much emphasis must be placed on efficient caching techniques for IoT microprocessors. 

Previous work has shown that specializing the cache configurations to different application or phase memory requirements can reduce the memory hierarchy's energy consumption by up to 62\% \cite{gordon-ross09}. In addition, our studies revealed that the cache is one of the more easily over-provisioned resources in an IoT microprocessor, resulting in high energy consumption with no performance benefits. The energy consumption can be quantifiably reduced, without any performance degradation, by dynamically changing the cache configurations (e.g., reducing the cache size). Thus, a prominent optimization for IoT microprocessors is dynamically configurable cache architectures that allow the cache’s parameter values to be specified/changed during runtime.

Three major challenges must be addressed in order to enable dynamically configurable caches for IoT microprocessors: \textit{augmenting caches for configurability}, \textit{cache tuning algorithms/heuristics}, and \textit{cache tuners}. In order to maximize the benefits of configurable caches, the required hardware optimizations to enable configurability must accrue minimal overhead. For example, a potential technique for enabling cache configurability uses bit-width configuration registers that allow a cache’s banks to be shutdown to configure the cache size, or concatenated to configure the cache associativity \cite{zhang03}. 

For optimal execution, the best configurations that achieve optimization goals and satisfy design constraints for different applications must be dynamically determined. Cache tuning determines the optimal cache configurations that match an application's runtime behavior. The cache tuning algorithm/heuristic can result in time and energy overheads, since the processor must stall during the tuning process. Much previous work (e.g., \cite{adegbija14_1,gordon-ross09,hajimiri12,zhang03,navarro15}) have proposed different algorithms/heuristics for cache tuning to minimize the potential of overhead and maximize the cache tuning benefits. These works offer a valuable foundation for IoT microprocessors. The intrinsic characteristics of the IoT necessitate further studies and development of innovative cache tuning techniques for IoT microprocessors that are low-overhead, robust, and versatile for the variety of applications that will execute on these microprocessors. 

To orchestrate the cache tuning process, hardware and/or software cache tuners employ cache tuning algorithms/heuristics to determine the best cache configurations to meet design constraints. However, the tuner could impose significant power, area, and/or performance overheads while exploring the configuration design space \cite{adegbija14_2}. Thus, to maximize the benefits of configurable caches in IoT microprocessors, novel cache tuners must be designed such that they constitute minimal overhead and effectively implement the cache tuning algorithms.

\subsection{Distributed Heterogeneous Architectures} 
Heterogeneous architectures \cite{kumar03} allow a coarse-grained specialization of system resources to application requirements by equipping a microprocessor with different kinds of cores or different core configurations. The different cores execute the same instruction set, but have different capabilities and performance levels. Thus, at runtime, the system software evaluates the resource requirements of applications or application phases and determines the core that best satisfies the optimization goals for the executing applications. 

One of the major advantages of heterogeneous architectures for IoT microprocessors, from a design perspective, is that existing cores (e.g., CPUs, DSPs, GPUs, etc.) can be reused in the implementation of heterogeneous microprocessors; this allows previous design and verification efforts to be amortized. However, unlike configurable architectures, heterogeneous architectures offer a much smaller design space, which necessitates greater design time effort in determining the best core configurations that will satisfy the application requirements. In addition, in a system with a large number of applications, heterogeneous cores may have a lower optimization potential than configurable cores, since there are fewer configurations to choose from in heterogeneous cores.

Much previous research efforts have targeted heterogeneous cores in general purpose computers, embedded systems, etc., but their applicability to IoT microprocessors have yet to be explicitly determined \cite{singh13,miorandi12}. Two major challenges that must be addressed in designing heterogeneous microprocessors for the IoT are the number and choice of cores, and scheduling of applications to the appropriate cores. In order to maximize the optimization potential, designers must expend a considerable amount of effort to determine the best cores or configurations to incorporate into the microprocessor. To provide an effective platform that satisfies the execution requirements of a wide variety of application characteristics, the selected cores must cater to a wide range of computational complexities and performance requirements. This effort would require a priori knowledge and analysis of the applications/application domains that will execute on the microprocessor. In addition, potential core configurations and their characteristics must be extensively analyzed.

Given the application execution requirements, the appropriate core on which to execute/schedule the application must also be determined either statically or dynamically \cite{arabnejad14}. Static scheduling suffices when the applications are known a priori. However, when the applications are unknown, dynamic scheduling evaluates application characteristics at runtime and schedules the applications to the appropriate cores. Much research is needed to develop low-overhead, computationally simple, and accurate scheduling techniques, for IoT microprocessors, that will achieve optimization goals and satisfy the microprocessors’ resource constraints.
\begin{figure}
	\centerline{\includegraphics[width=0.4\textwidth]{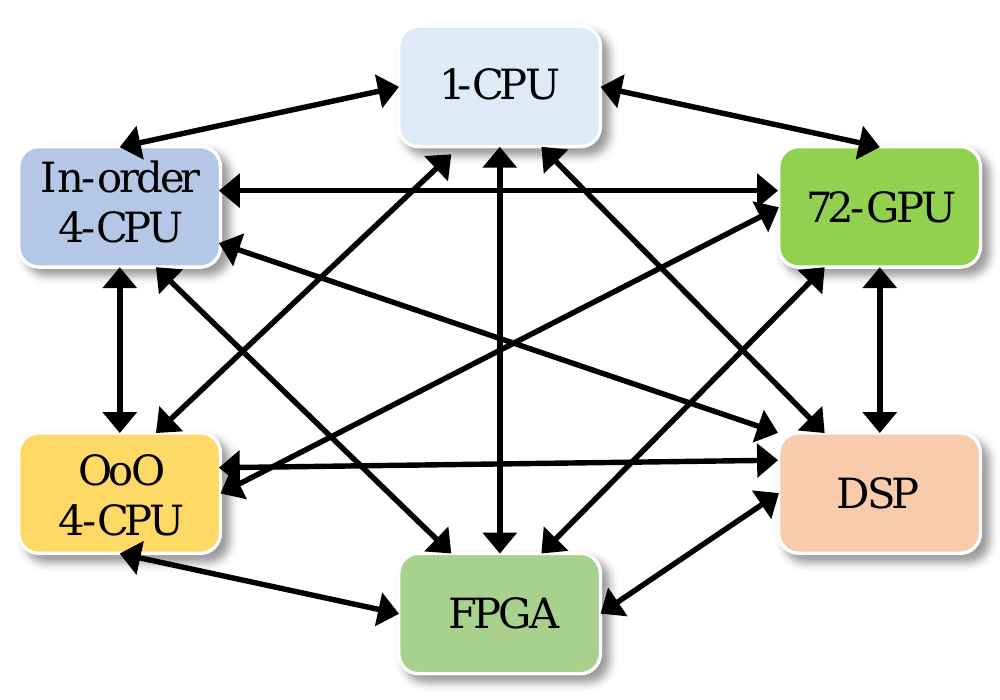}}
	\caption{Distributed heterogeneous architectures.}
	\label{fig:heterogeneous}
\end{figure}

An alternative to heterogeneous cores on a single device is a network of distributed heterogeneous architectures. Figure \ref{fig:heterogeneous} illustrates the distributed heterogeneous network. Different devices are equipped with different computational resources that may be required by the devices at different times. For example, Figure \ref{fig:heterogeneous} depicts six nodes containing six different kinds of microprocessors: a single core in-order microprocessor (1-CPU), a 72-core general purpose GPU (72-GPU), a DSP, an FPGA, a quad-core out-of-order processor (OoO 4-CPU), and a quad-core in-order processor (In-order 4-CPU). Assuming an 8-threaded application \textit{A} with deadline constraints arrives on 1-CPU---1-CPU may be under-provisioned for \textit{A}'s execution (i.e., the execution time on 1-CPU will exceed the deadline)---an alternative node in the network can be used to execute \textit{A}. Apart from determining which node contains sufficient resources to execute \textit{A}, the costs---energy and time---to transfer \textit{A} from 1-CPU to the right-provisioned node, in addition to the time to retrieve the results, must not negate the savings from executing \textit{A} on a right-provisioned core

To enable distributed heterogeneous architectures, several research challenges must be addressed: benefits of waiting for a right-provisioned node if the node is busy; tradeoffs of time---execution and transmission times---and energy in the presence or absence of deadline constraints; runtime, low-overhead determination of right-provisioned nodes; developing portable applications and standardizing the communication protocols between different kinds of nodes.

\begin{table*} 
	\begin{center}
		\captionof{table}{Summary of computing paradigms and potential benefits}
		\begin{tabular}{| l || l | l |} 
			\hline \label{table:paradigms}
			\textbf{Computing Paradigm} & \textbf{Benefits} & \textbf{References}\\ \hline
			Configurable architectures & Efficiency, future-proof, configurability & \cite{folegnani01,kora13,abella03,efthymiou03,zhang03,gordon-ross09,hajimiri12,adegbija14_1,adegbija14_2,navarro15}\\ \hline
			Distributed heterogeneous architectures & Extensibility, efficiency, future-proof & \cite{kumar03,singh13,miorandi12} \\ \hline
			Energy harvesting and non-volatile processors & Efficiency & \makecell[l]{ \cite{chandrakasan08,gudan12,gollakota14,raghunathan05,li11_solar,visser08,liu15_ambient}, \\ \cite{wang12,yu11,liu16,zwerg11,choi13,swaminathan11,onkaraiah12,ma15}} \\ \hline
			Approximate computing & Efficiency & \makecell[l]{\cite{chippa13,kugler15,samie15,kahng12,venkataraman15,han13,lu04,du12}, \\ \cite{kulkarni11,kyaw10,shin10,shin11,sampson14,bortolotti14}} \\ \hline
			In-memory processing & Efficiency, extensibility & \cite{zaharia12,jiang14,cruz12,kang16,kang14,kang15,zhang16,ahn15} \\ \hline
			Secure microarchitectures & Security & \cite{koeberl14,kocher04,sadeghi15,kanuparthi13,kermani13,serpanos13,babar11,crenne13,zhuang14,dai17}\\ \hline
			
		\end{tabular}
	\end{center}
\end{table*}

\subsection{Energy Harvesting and Non-Volatile Processors}

One of the most important, and most persistent, challenges for IoT devices is energy consumption. Energy harvesting is a promising technique for replacing or supplementing energy sources (e.g., batteries), especially in ultra-low power applications. Energy can be harvested from several sources, including solar, thermal gradients, radio frequency radiation, etc. \cite{chandrakasan08,gudan12,gollakota14,raghunathan05,li11_solar}. These energy sources, however, are typically not reliable; external factors---distance from a power source, physical obstacles, electromagnetic signals---can disrupt the energy supply \cite{visser08}. Due to this unreliability, traditional processors may be impractical for systems equipped with energy harvesting---when the energy supply is disrupted, volatile processors will lose their operating state.

Non-volatile processors \cite{liu15_ambient} use non-volatile storage components---non-volatile memories---to store the processor state when the power supply is disrupted. When the power is restored, the processor's state is restored from the the non-volatile memory to continue execution. Thus, non-volatile processors allow continuous computation despite power disruptions. For example, one of the earlier non-volatile processors \cite{wang12} allowed system states to be backed up within 7$\mu$s and restored within 3$\mu$s. Additionally, since embedded systems typically spend a significant amount of time idling, resulting in high leakage power, non-volatile processors can reduce the idle power by allowing the system to be shut down while idle. The state can then be instantaneously restored on wake-up \cite{yu11,liu16}.

There are several potential non-volatile architectures that can be used at different abstraction levels to achieve non-volatile processors for energy harvesting systems. For example, various FeRAMs, STT-RAMs, PCRAMs, ReRAMs have been explored for use in energy harvesting systems \cite{zwerg11,choi13,swaminathan11,onkaraiah12,liu15_ambient}. Several factors must be considered when selecting an energy harvesting non-volatile processor system. The input power characteristics---for example, the power behavior when interrupted---affect the choice of non-volatile architectures \cite{ma15}. The application characteristics also affect the choice of non-volatile architectures. Application characteristics, such as deadlines, real-time, and quality of service (QoS) requirements, must also be taken into consideration. For example, solar powered systems can typically be used to meet real-time QoS requirements more effectively than RF or thermal source \cite{ma15}. Much research is required to quantify the tradeoffs of different energy sources with respect to the non-volatile architectures.

\subsection{Approximate Computing}

\textit{Approximate computing} has recently gained a lot of traction as a viable alternative to exact computing. Exact computing targets exact numerical or Boolean equivalence, while approximate computing allows a non-exact, inaccurate result that maintains the desired output quality \cite{chippa13}. Approximate computing allows new optimization options for processors that execute \textit{resilient applications}---applications that can produce outputs of sufficient quality despite some imprecise computations, e.g., signal processing, multimedia, graphics, etc. Allowing bounded approximation in processors can provide significant performance and energy gains, while achieving an acceptable amount of accuracy.

Approximate computing can enable energy-efficient edge computing in IoT devices, such as wearable electronics \cite{kugler15,samie15}. One of the first steps to incorporating approximate computing into IoT devices is identifying the devices' applications, and the applications' resilience to computing error. In \cite{chippa13}, Chippa et al. presented an \textit{automatic resilience characterization} framework to evaluate how amenable an application is to approximate computing. This framework uses approximation models that evaluate different approximate computing techniques for different partitions of an application. This framework relies on significant amounts of a priori knowledge about the executing applications, such as, the computational patterns and input data. Since applications typically have different execution phases, and the phase behaviors could change at runtime, key to efficient approximation is a runtime framework that automatically detects resilient application phases and adjusts the computing exactness to match the currently executing phase's resilience \cite{kahng12}.

Several optimizations to enable approximate computing have been developed at different abstraction levels. Using various digital signal processing filters and an electrocardiogram (ECG) application, Venkataraman et al. \cite{venkataraman15} presented a system-level design flow to study a system's exactness and used elimination heuristics to explore the design space under inexactness, area, and energy constraints. Several other techniques have been proposed \cite{han13} for achieving inexactness at the circuit level through different approximate components, such as approximate adders \cite{lu04,du12}, approximate multipliers \cite{kulkarni11,kyaw10}, and approximate logic synthesis \cite{shin10,shin11}. Similarly, approximate computing can also be exploited at the memory level, for example, by storing data approximately \cite{sampson14} or through systems that can tolerate memory errors while maintaining the desired quality of service \cite{bortolotti14}. 

\subsection{In-Memory Processing}

In-memory processing---or \textit{processing in memory}---has been studied in relation to big data and distributed computing systems \cite{zaharia12,jiang14}. In-memory processing addresses the well-known processor-memory performance gap through internal memory accesses; it avoids delays caused by off-chip communication. Caches have been widely used to bridge the processor-memory performance gap; in-memory processing further reduces this gap by allowing computations to be performed \textit{on} the memory chip without the need for processor-to-memory communication. With the growth of the IoT, massive amounts of data generated, and resource constraints of IoT devices, in-memory processing offers an attractive optimization for IoT devices.

One of the major attractions of in-memory processing for IoT devices, in the context of edge computing, is the need for real-time in-situ data processing on large data volumes. The data processing must be energy-efficient and involve low hardware overhead. To achieve such capabilities, researchers have explored inherently robust brain-inspired models of computation that involve highly efficient inference applications \cite{cruz12}. An example of such a computing model is the \textit{sparse distributed memory (SDM)}, which can be trained to remember sparse data vectors and retrieve them when presented with noisy or incomplete versions of the vectors \cite{kanerva88}. However, SDM architectures are challenging due to the often conflicting design goals of achieving both high throughput and energy efficiency.

To enable sparse distributed memory, \textit{compute memory} has been proposed as a viable implementation architecture \cite{kang16}. Compute memory \cite{kang14,kang15} is an in-memory processing architecture that implements both memory and processing in a single architecture in order to completely eliminate the processor-memory interface. The compute memory architecture implements inference algorithms in the periphery of the memory array, and does not modify the core bit-cell array, thus maintaining the storage density. The compute memory is able to implement operations, such as the sum of absolute differences, signed multiplication, etc. The SDM implementation, using compute memory, has been shown to achieve both high throughput and energy efficiency for data-rich applications, such as pattern recognition \cite{kang16}. 

However, most current compute memory implementations are application-specific. Ahn et al. \cite{ahn15} proposed a processor in memory application that uses specialized instructions, called PIM-enabled instructions, to invoke in-memory computations. The goal of the proposed work was to allow processing in memory operations to be compatible with existing systems and applications, without the need to specifically design the processor in memory for specific applications. Much work exists to extend compute memory architectures to multi-application use-cases, with minimal overheads.

Another architecture, similar to the compute memory, with high potential for IoT devices is the \textit{compute sensor} \cite{zhang16}, which offers in-sensor processing. The compute sensor takes advantage of machine learning algorithms' inherent adaptability to noise, and embeds information processing functionality for these algorithms into the sensor substrates. The compute sensor eliminates the sensor-processor interface, wherein sensed data is typically transmitted to a processor for data visualization, as is the case in traditional sensors. The compute sensor significantly reduces energy consumption and latency of feature extraction and classification functions, without sacrificing accuracy \cite{zhang16}.

\subsection{Secure Microarchitectures}

Security applications are some of the most important applications that will be executed on IoT microprocessors. The IoT's characteristics---pervasiveness, interdependence, connectedness, mobility---makes IoT devices inherently vulnerable to increasing number of attacks. IoT devices are susceptible to physical, side channel, cryptoanalysis, software, and network attacks. Security can no longer be an afterthought in microprocessor design; it is imperative that microprocessors are designed to be inherently secure. However, security in IoT devices is especially challenging due to the typically stringent resource constraints of these devices. 

Most IoT devices will have no hardware support for virtualization or enhanced security features, such as trusted execution \cite{koeberl14}. The processing capabilities of IoT devices' microprocessors may be exceeded by the resource requirements of security processes and algorithms. As a result, security designers often need to trade off other vital optimization goals, such as energy, performance, or cost \cite{kocher04}. Apart from the resource constraints of IoT device microprocessors, these devices also generate massive amounts of data, some of which may contain private or sensitive information \cite{sadeghi15}.

Much previous research has proposed hardware security techniques for embedded systems \cite{kanuparthi13,kermani13,serpanos13,babar11}. While most of these techniques can also be employed in IoT devices, one critical requirement for IoT devices is the need to have runtime configurable hardware security policies that can adapt to varying security requirements \cite{crenne13}. This requirement is motivated by the resource constraints of IoT devices and the fact that the sensitivity of various computations vary depending on the executing applications. Thus, secure microarchitectures for the IoT must have the capability to be adjusted to satisfy specific applications' or tasks' needs, while incurring minimal overheads.

There are currently very few techniques that have been proposed for configurable hardware security in microprocessor architectures, especially for IoT devices \cite{crenne13}. The main advantage of security at the level of the microarchitecture is that microarchitectures can typically be easily augmented for runtime configurability. In addition, configurable microarchitectures can be used to achieve multiple optimization goals. Thus, ensuring hardware security, using configurable microarchitectures, need not be at the expense of other optimization goals, such as energy consumption. For example, configurable caches can be used as a moving target defense \cite{zhuang14} against side-channel attacks in caches, while also maintaining the other optimization benefits of configurability (e.g., energy and performance) \cite{dai17}. 

\section{Conclusions}
The Internet of Things (IoT) is expected to transform life, business, and the global economy. The IoT's scale and rapid proliferation will generate massive amounts of data that will result in communication bandwidth bottlenecks, and latency and energy overheads. Edge computing significantly reduces these overheads by equipping IoT devices with right-provisioned microprocessors and algorithms that can perform computations on the edge nodes to interpret, visualize, and use data.

This paper presented an overview of microprocessor characteristics that will support the growth of the IoT, from an edge computing perspective, and optimizations that will enable those characteristics. The survey presented herein should provide researchers with a foundation for designing IoT microprocessors that are efficient, configurable, secure, future-proof, and extensible. We have also discussed some of the challenges with achieving the discussed optimizations, and presented some potential solutions for addressing the challenges. Since edge computing on the IoT is a growing area of research, this study provides a foundation for further research into application requirements and microprocessor optimizations that will support edge computing in next-generation IoT devices.

\ifCLASSOPTIONcaptionsoff
  \newpage
\fi



%

\bibliographystyle{abbrv}
\bibliography{../../References/tosiReferences}

\vspace{-20pt}
\begin{IEEEbiography}[{\includegraphics[width=1in,height=1.25in,clip,keepaspectratio]{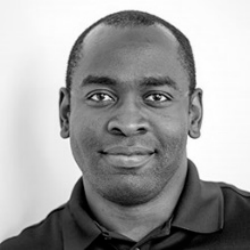}}]{Tosiron Adegbija}
received his M.S and Ph.D in Electrical and Computer Engineering from the University of Florida in 2011 and 2015, respectively and his B.Eng in Electrical Engineering from the University of Ilorin, Nigeria in 2005. He is currently an Assistant Professor of Electrical and Computer Engineering at the University of Arizona, USA. His research interests are in computer architecture, with emphasis on adaptable computing, low-power embedded systems design and optimization methodologies, and microprocessor optimizations for the Internet of Things (IoT). He received the best paper award at the Ph.D forum of IEEE Computer Society Annual Symposium on VLSI in 2014.
\end{IEEEbiography}
\begin{IEEEbiography}[{\includegraphics[width=1in,height=1.25in,clip,keepaspectratio]{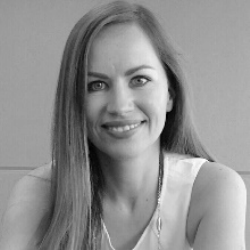}}]{Anita Rogacs}
received her M.S. and Ph.D in Mechanical Engineering from Stanford University and her B.S. in Mechanical Engineering from San Jose State University. She is recipient of National Science Foundation and Sandia National Laboratories Research Fellowships. At HP Labs she built Life Sciences Labs, which is now home to research efforts in areas of plasmonics, Raman spectroscopy, microfluidics, analytical chemistry, molecular biology and data mining. She is also the HP Labs Principle Investigator of the CRADA collaboration with the Federal Food and Drug Administration, which aims to develop screening methods using Surface Enhanced Raman Spectroscopy (SERS) sensors.
\end{IEEEbiography}

\begin{IEEEbiography}[{\includegraphics[width=1in,height=1.25in,clip,keepaspectratio]{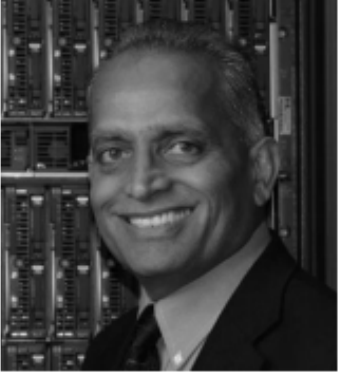}}]{Chandrakant Patel}
 is currently Chief Engineer and Senior Fellow of HP Inc. Patel has led HP Labs in delivering innovations in chips, systems, data centers, storage, networking, print engines and software platforms. He is a pioneer in thermal and energy management in data centers, and in the application of information technology for available energy management at the scale of cities. Patel is an ASME and an IEEE Fellow, and has been granted 151 patents and published more than 150 papers. An advocate of a return to fundamentals, he has served as an adjunct faculty member in engineering at Chabot College, U.C. Berkeley Extension, San Jose State University and Santa Clara University. In 2014, Patel was elected to the Silicon Valley Engineering Hall of Fame.
\end{IEEEbiography}
\begin{IEEEbiography}[{\includegraphics[width=1in,height=1.25in,clip,keepaspectratio]{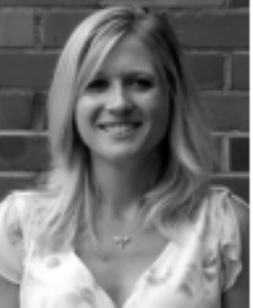}}]{Ann Gordon-Ross}
received the B.S. and Ph.D. degrees in computer science and engineering from the University of California, Riverside, USA, in 2000 and 2007, respectively. She is currently an Associate Professor of Electrical and Computer Engineering with the University of Florida, USA, and a member of the NSF Center for High Performance Reconfigurable Computing with the University of Florida. She is also a Faculty Advisor of the Women in Electrical and Computer Engineering and the Phi Sigma Rho National Society for Women in Engineering and Engineering Technology, and an active member of the Women in Engineering Proactive Network. Her research interests include embedded systems, computer architecture, low-power design, reconfigurable computing, dynamic optimizations, hardware design, real-time systems, and multicore platforms. She was a recipient of the CAREER Award from the National Science Foundation in 2010, best paper awards at the Great Lakes Symposium on VLSI in 2010 and the IARIA International Conference on Mobile Ubiquitous Computing, Systems, Services and Technologies in 2010, and the Best Ph.D. Poster at IEEE Computer Society Annual Symposium on VLSI in 2014. She is very active in promoting diversity in STEM fields, and has been a Guest Speaker at several international workshops/conferences on this topic, organizes workshops, and participates in local outreach programs at local K-12 schools.
\end{IEEEbiography}



\balance

\end{document}